# Ultrahigh Precision Absolute and Relative Rotation Sensing using Fast and Slow Light

M.S. Shahriar, G.S. Pati, R. Tripathi, V. Gopal, M. Salit and K. Salit

*Department of Electrical Engineering and Computer Science*

*Northwestern University, Evanston, IL 60208*

**Abstract**

We describe a resonator based optical gyroscope whose sensitivity for measuring absolute rotation is enhanced via use of the anomalous dispersion characteristic of superluminal light propagation. The enhancement is given by the inverse of the group index, saturating to a bound determined by the group velocity dispersion. We also show how the offsetting effect of the concomitant broadening of the resonator linewidth may be circumvented by using an active cavity. For realistic conditions, the enhancement factor is as high as $10^6$. We also show how normal dispersion used for slow light can enhance relative rotation sensing in a specially designed Sagnac interferometer, with the enhancement given by the slowing factor.

PACS numbers: 42.5, 42.50.Ct, 42.50.Gy, 42.60.Da

# 1. Introduction

Recent experiments have demonstrated very slow as well as superluminal group velocities of light using many different processes[1,2,3,4,5]. Extreme dispersion in a slow-light medium has recently been considered to boost the rotational sensitivity of optical gyroscopes via the relativistic Sagnac effect[6]. However, the mechanism by which strong material dispersion affects the Sagnac effect has not been analyzed extensively. It is commonly perceived that the Sagnac time delay or phase shift in the presence of a dielectric medium is independent of the refractive property of the medium[7,8]. Although this is true for a co-moving medium and absolute inertial rotation, it is not true if the medium is moving with respect to the inertial frame of rotation. Recently, Leonhardt and Piwnicki[6] proposed a slow-light enhanced gyroscope; however, they did not point out that this enhancement can not be achieved when measuring absolute rotation. Later, Zimmer and Fleischhauer[9] refuted, incorrectly, the idea of any slow light based sensitivity enhancement in rotation sensing, and proposed an alternative hybrid light and matter-wave interferometer, based on slow light momentum transfer from light to atoms.

Matsko et al.[10] proposed a model where they consider the use of microresonators instead of atoms to provide the positive dispersion. The analysis in this paper makes use of the classical (non-relativistic) model, which yields results that are incorrect when a material medium of non-unity index is present. The authors published an erratum[11] in which they point out this mistake[12]. However, the erratum still claims that the rotation sensitivity is enhanced by the group index. This conclusion is correct, although potentially misleading. When a single loop used in an interferometric Sagnac rotation sensor is replaced by a cavity, the sensitivity is enhanced by a factor equaling its finesse. This factor also equals the ratio of the phase velocity to the group velocity in the cavity. Therefore, it can be represented as an effective group index, or, equivalently, a slowing factor. Physically, this enhancement results from the fact that the counter-propagating beams travel a number --- which equals the finesse --- of times before interfering with each other. Thus, this result makes no statement about the effect of positive dispersion in the propagation medium on the sensitivity of Sagnac effect based rotation sensing[13]. Essentially the same conclusion applies to the micro-resonator based rotation sensing architecture proposed by Scheuer et al[14].

In this paper, we first present a systematic but brief review of the role of positive dispersion in the propagation medium in measuring rotation via the Sagnac effect. This allows us to put the various claims regarding potential enhancement of the sensitivity of a Sagnac gyroscope using the slow-light effect in clear perspective. Specifically, we establish that there are no known configurations under which a positive dispersion in the propagation medium can enhance the sensitivity of measuring absolute rotation. We then identify an explicit configuration where such a positive dispersion leads to a strong enhancement (by a factor matching the group index) in measuring *relative* rotation between the propagation medium and the rest of the gyroscope. This is a manifestation of the fact that in an interferometric optical gyroscope that allows relative motion between the medium and the interferometer, the dispersive drag coefficient[8,15] can influence the Sagnac fringe shift.

For most practical applications such as inertial navigation and fundamental studies in astrophysics (e.g., precision measurement of the Lens-Thirring rotation) and geology, the measurement of relative rotation is of no use. What is needed for these applications is an ability to measure the absolute rate of rotation of the whole gyroscope, including the propagation medium. As the primary contribution of this paper, we show here that dispersion in the propagation medium can indeed be used to enhance the sensitivity of measuring *absolute rotation*. Specifically, we show that the enhancement occurs if the dispersion is anomalous, characteristic of superluminal light propagation, in a resonator based gyroscope where the Sagnac effect manifests itself in the form of the frequency splitting of cavity modes. Unlike an interferometer, frequency-splitting in a resonator gyroscope is governed by optical path lengths that are always dependent on the refractive property of the medium[15,16,17]. We discuss the effect of dispersion on such a resonator-based gyroscope, and show that the rotation sensitivity can be increased by as much as $10^6$ by using what we call critically anomalous dispersion (CAD), corresponding to the group index approaching a null value. Such a sensor can be implemented using bi-frequency pumped Raman gain in an atomic medium[18], for example. We also show how the offsetting effect of the concomitant broadening of the resonator linewidth can be circumvented by using an active cavity.

The rest of the paper is organized as follows. In Section 2, we present a brief summary of the role of medium dispersion in the Sagnac effect, using the axioms of Special Relativity. We establish that there are no known configurations under which a positive medium dispersion can enhance the sensitivity of measuring absolute rotation, and identify an explicit configuration where such a positive medium dispersion leads to enhancement in measuring *relative* rotation. The published claims regarding the use of such a dispersion in enhancing the sensitivity of a gyroscope are then summarized and critiqued in this context. In Section 3, we present the key idea of this paper, by describing how a resonator based Sagnac gyroscope with an intra-cavity negative dispersion (i.e., fast-light) medium near the CAD condition leads to an strong enhancement in the sensitivity of measuring absolute rotation. In Section 4, we establish the effective equivalence between rotation sensing and measurement of mirror displacement in a ring resonator, and use this model to present a simple and transparent interpretation of the fast-light enhanced rotation sensing. This model is also used to show how this enhancement is naturally accompanied by the so-called White Light Cavity (WLC) effect whereby the linewidth of the cavity is broadened without affecting its storage time. In Section 5, we show how, in a passive ring resonator gyroscope loaded with a negative dispersion medium near the CAD condition, the enhancement in sensitivity to rotation is counterbalanced by the concomitant linewidth broadening (i.e., the WLC effect), so that there is virtually no improvement in the minimum measurable rotation rate. In Section 6, we show how this offsetting effect of linewidth broadening can be overcome by using a fast-light enhanced ring laser gyroscope. In Section 7, we outline an explicit scheme for realizing fast-light enhanced gyroscope, and present numerical estimates for the prospect of measuring the Lens-Thirring rotation using such a gyroscope. We also refer briefly to recent experimental work we have done to establish key aspects of this gyroscope. Finally, we conclude with a summary and acknowledgments in Section 8.

## 2. Enhancement of the Sagnac Effect in the Presence of Positive Dispersion in the Propagation Medium

Let us consider first waves in a Mach-Zehnder interferometer (MZI), as shown in Fig. 1, that are constrained to propagate in a circular path (for convenience in analysis) with a radius R (the effect occurs for paths that are rectilinear as well, as shown in the diagram). In general, the wave could be of any kind: optical, matter or acoustic. For the clockwise CW(+) and counter-clockwise CCW(-) directions, the relativistic velocities $V_R^\pm$ of the phase fronts (PFs), the time $T^\pm$ of PFs to travel from BS1 to BS2, and the effective distances $L^\pm$ from the BS1 to BS2 are related as follows

$$V_R^\pm = \frac{V_P \pm v}{1 \pm V_P v / C_o^2}, \quad L^\pm = \pi R \pm v T^\pm, \quad T^\pm = \frac{L^\pm}{V_R^\pm} \qquad (1)$$

where $V_P$ is the velocity of each PF in the absence of rotation, $C_o$ is the velocity of light in vacuum, $v=\Omega R$ is the tangential velocity of rotation, and $\Omega$ is the rotation rate.

Thus, the time delay between the propagating waves and their relative phase shift at the detector are given by

$$\Delta t_o = T^+ - T^- = \pi R \left[ \frac{(V_R^- - V_R^+) + 2v}{(V_R^- + v)(V_R^+ - v)} \right] = \frac{2 A \Omega}{C_o^2 (1 - \beta^2)} \approx \frac{2 A \Omega}{C_o^2}, \quad \beta \left( \equiv \frac{v}{C_o} \right) << 1, \qquad (2)$$

$$\Delta \phi_o = \omega \Delta t_o = \frac{4 \pi A \Omega}{\lambda_o C_o}$$

where $A=\pi R^2$ is the enclosed area, $\omega$ is the angular frequency, $\lambda_o$ is the vacuum wavelength of the wave and $\beta<<1$ is the boost parameter. The form of the time delay in eqn. 2 due to the Sagnac effect attests to the fact that the time delay is simply a geometric effect, attributable to relativistic time dilations, and does not depend at all on the velocity of the wave. It only involves the free space velocity of light, even if acoustic waves or matter waves are used. The familiar result of matter-wave Sagnac phase shift, $\Delta\phi = 4\pi m A\Omega/h$, can be obtained from eqn. 2 by inserting the Compton frequency $\omega = (mc_o^2/\hbar)$. Note that the Sagnac effect can be explained fully within SR, without invoking general relativity, if the degree of space-time curvature is insignificant, which is the case for optical loops typically used in Sagnac interferometers[8].

For the case of optical waves in a Sagnac interferometer consisting of a refractive medium, the phase shift $\Delta\phi_o$ in eqn. 2 is independent of the refractive index n. This result is somewhat counterintuitive, since it is expected that the light beams experience index-dependent drag in the moving medium. Thus the travel time delay around the optical circuit could be expected to depend on n. However, this is not true for a co-rotating medium in the Sagnac interferometer. One can explicitly show this by re-expressing the velocities $V_R^\pm$ of the CW and CCW phase fronts in terms of the Fresnel drag:

$$V_R^\pm = \frac{C_o}{n} \pm v\,\alpha_F, \quad \alpha_F = (1 - \frac{1}{n^2}) \qquad (3)$$

where $C_o/n$ is the velocity $V_p$ of the phase fronts in the absence of rotation, and the term '$\alpha_F$' is the Fresnel drag coefficient. Using eqn. 2, one can write the time delay and the phase shift in a generalized form

$$\Delta t \approx \left(n^2(1-\alpha_F)\right)\Delta t_o, \quad \Delta\phi = \left(n^2(1-\alpha_F)\right)\Delta\phi_o \qquad (4)$$

Substituting '$\alpha_F$' from eqn. 3 in eqn. 4, one gets the same results as in eqn. 2. This is because the derivation using SR already includes the effect of the medium motion on light propagation. However, the magnitude of this effect is independent of the Fresnel drag coefficient, as well as the refractive index, for a Sagnac interferometer that uses a common frame of rotation for the light source, the interferometer, and the medium. Such a system does not possess (to first-order) Doppler shift of any kind between the medium and the source or the interferometer, since they are all co-rotating at the same rate $\Omega$.

One can relax this constraint and allow several modifications of the Sagnac interferometer where relative motion between the frame and the medium introduces a Doppler shift. As we will show, under this condition, light drag resulting from dispersion affects the final fringe shift. Such a system, however, can only be used in relative rotation sensing. Fig.2A shows another MZI considered for Sagnac interferometry, allowing relative motion of the medium with respect to the interferometer. Note first that in this model the connection with the first BS is established using a flexible fiber, thus making it possible to make the MZI rotate without rotating the source. If the source is stationary while the frame and the medium co-rotate, a Doppler shift is produced at the first BS. However, this shift does not affect the phase shift, since the BS effectively serves as the source for both arms. Therefore, the fringe shift described in eqn. 2 still holds. Next, consider the case where the medium inside the MZI paths is allowed to move at a velocity $V_M$ with respect to the MZI frame (i.e. the mirrors and beam splitters). As seen by medium, the CW(+) and CCW(-) beams are now Doppler shifted by equal and opposite amounts, given by $\Delta\omega^\pm = \pm\omega V_M / C_o$. Expressions for relativistic velocities can be given by including first-order dispersion in n(ω) as

$$V_R^\pm = \frac{C_o}{n_o}(1 - \frac{\Delta\omega^\pm}{n_o}\frac{\partial n}{\partial \omega}) \mp v\alpha_F = \frac{C_o}{n_o} \mp \frac{V_M}{n_o^2}\omega\frac{\partial n}{\partial \omega} \mp v\alpha_F = \frac{C_o}{n_o} \mp V_M\frac{(n_g - n_o)}{n_o^2} \mp v\alpha_F \quad (5)$$

where $n_o$ is the phase index at the input frequency of ω, $\alpha_F = (1 - 1/n_o^2)$ is the Fresnel drag coefficient, and $n_g (= n_o + \omega\partial n/\partial\omega)$ is the group index, defined as the ratio of the free space velocity of light to the group velocity at ω. As a specific example, consider a situation where the medium is non-comoving, as illustrated in Fig. 2B. and 2C., giving rise to $V_M \cong (-v)$. The relativistic velocities in this case are given by

$$V_R^\pm \approx \frac{C_o}{n_o} \pm v\alpha_L, \quad \alpha_L = \left[1 - \frac{1}{n_o^2} + \frac{(n_g - n_o)}{n_o^2}\right] \quad (6)$$

where $\alpha_L$ is the so-called Laub drag coefficient[19]. The expressions for time delay and phase shift are obtained as $\Delta t \approx (n^2(1-\alpha_L))\Delta t_o$, $\Delta\phi \approx (n^2(1-\alpha_L))\Delta\phi_o$.

Unlike the Doppler-shift free situation, the coefficients of time delay and phase shift now depend on the refractive property of the medium. Of course, in the absence of dispersion (i.e., $n_g=n_o$), one recovers the same results as in eqn. 2. In the presence of a highly dispersive medium with $n_g \gg n_o$ (i.e. $\partial n/\partial\omega \gg n_o/\omega$), characteristic of extreme normal dispersion in a medium that produces slow-light, the time delay and the phase shift are given by $\Delta t \approx n_g \Delta t_o$, $\Delta\phi \approx n_g \Delta\phi_o$.

Thus, the rotation sensitivity of a slow-light based gyroscope scales directly with the group index $n_g$. Experimental observation of slow light reported in atomic and solid media using EIT[1,2] have shown group index $n_g$ to be as large as $10^8$. Thus one can conceivably design an ultra-precise rotation sensor for sensing relative rotation of a local system using a uniformly spinning slow light medium in the interferometric beam path. Finally, we note that the experiment outlined here for detecting the relative rotation between a dispersive medium and the rest of the gyroscope is fundamentally equivalent to the Fizeau experiment[20].

We now summarize briefly some of the ideas that have been presented in the literature regarding the use of slow-light for enhancement of rotation sensing. First, not that the result we presented above for using slow-light to enhance the sensitivity of measuring relative rotation is essentially equivalent to what was presented in ref. 6. However, as we noted earlier, ref. 6 does not point out that this enhancement can not be achieved when measuring absolute rotation. This result also shows the claim by Zimmer and Fleischhauer that slow-light can not enhance the sensitivity of rotation sensing to be inaccurate. Finally, as discussed in the introduction, the results presented by Matsko et al. does not deal with any potential enhancement of rotation sensing via the use of positive dispersion in the propagation medium. Rather, it is essentially a restatement of the well-established fact that when a single loop used in an interferometric Sagnac rotation sensor is replaced by a cavity, the sensitivity is enhanced by a factor equaling its finesse. Essentially the same conclusion applies to the micro-resonator based rotation sensing architecture proposed by Scheuer et al.

## 3. Enhancement of the Sagnac Effect in a Ring Resonator in the Presence of Negative Dispersion

As mentioned earlier, one of the drawbacks of the slow-light enhanced interferometric configuration above is that it can not be employed in sensing absolute rotation. This constraint can be overcome by considering the Sagnac effect in a ring cavity or resonator. Unlike the interferometer where the frequency of light remains unchanged by rotation, the resonant frequencies of the cavity modes in a resonator get modified as a result of rotation. Figure 3 shows the schematic of a passive ring cavity (PRC) along with the necessary detection mode and servo mechanism to be used in a gyroscope. For analytic convenience, we assume the light path in the resonator to be circular with a radius R and an enclosed area A (the effect occurs even when the paths are rectilinear, as shown in the diagram). In the absence of rotation, the light in the cavity resonates in both directions with a frequency $\omega_o$ given by $\omega_o = (C_o/n_o)(2\pi N/P)$, where N is an integer, and $P = 2\pi R$ is the perimeter. In the presence of rotation, the resonance frequencies are different for the CW(+) and CCW(-) directions, and are given by:

$$\omega^{\pm} \cong V_E^{\pm} \cdot \frac{2\pi N}{P}; \quad V_E^{\pm} = V_R^{\pm} \mp v; \quad \delta\omega_o \left(\equiv \omega^- - \omega^+\right) = \frac{\omega_o}{C_o n_o}(2\Omega R) = \frac{\omega_o}{C_o n_o} \cdot \frac{4\Omega A}{P} \qquad (7)$$

where $V_R^{\pm}$ are the relativistic velocities given in eqn. 3, $V_E^{\pm}$ are the effective velocities and $A = \pi R^2$ is the area enclosed by the resonator. The frequency difference $\Delta\omega_o$ which is proportional to the rotation rate $\Omega$, can be experimentally measured from heterodyne beat note of transmitted signals by adjusting the frequency of AOM1(AOM2) to maximize the cavity transmission in CCW(CW) direction. *It is important to emphasize that the appearance of 1/n$_o$ in the expression for δω$_o$ in eqn. 7 results from the combination of the facts that* $(1-\alpha_F) = 1/n_o^2$, *and that, in the absence of rotation, the effective velocity of light in the medium is* $C_o/n_o$.

Note that even though we have calculated the difference between the resonant frequencies in each direction, we could also simply calculate the frequency offset in each direction independently. Let us define $\Delta\omega_o^- = (\omega^- - \omega_o)$ and $\Delta\omega_o^+ = (\omega^+ - \omega_o)$, so that $\delta\omega_o = (\Delta\omega^- - \Delta\omega^+)$. Using the expressions for $\omega^\pm$ from eqn. 7, we can show that

$$\Delta\omega_o^- = \frac{\omega_o}{C_o n_o} \cdot \frac{2\Omega A}{P}; \quad \Delta\omega_o^+ = -\frac{\omega_o}{C_o n_o} \cdot \frac{2\Omega A}{P} \tag{8}$$

Therefore, for the Passive Ring Cavity Gyroscope (PRCG), one can measure the rotation rate by monitoring the shift in the resonance frequency for one direction (CW or CCW) alone. Experimentally, both directions are measured in order primarily to eliminate common-mode effects. The same is true for the Ring Laser Gyroscope (RLG) as well.

Thus, in measuring absolute rotation, the PRCG (as well as the RLG) differ fundamentally from the MZI based gyroscope (MZIG). For the MZIG, one must have beams propagating in both directions, and the Sagnac effect is manifested only as the difference between the phase shifts due to a rotation. In contrast, for the PRCG (and the RLG), the Sagnac effect is manifested through the actual phase shift in either direction (CW or CCW). This is partly the reason why for the MZIG case, the measurement signal is independent of the index, while for the PRCG (and the RLG) it depends explicitly on the index.

In this derivation, we have implicitly assumed the refractive index to be independent of frequency (i.e. no dispersion). When the effect of dispersion is taken into account, the result changes significantly. Without any loss of generality, one can write

$$\omega^\pm = \omega_o \pm \delta\omega/2 = V_E^\pm \cdot (2\pi N/P) \tag{9}$$

where δω is a parameter whose amplitude is to be determined. The effective velocities, $V_E^\pm$ can be written as

$$V_E^\pm = V_R^\pm \mp v = \frac{C_o}{n(\omega^\pm)} \cdot \left[1 \pm \frac{v}{C_o n(\omega^\pm)}\right] \tag{10}$$

Expanding the value of n(ω) around $n_o$, we get

$$V_E^\pm = \frac{C_o}{n_o} \cdot \left[ 1 \pm \frac{v}{C_o n_o} \mp n' \frac{\delta\omega}{2} \right], \quad n' \equiv [\partial n/\partial \omega]/n_o \qquad (11)$$

Substituting eqn. 11 in eqn. 9, one gets a self-consistent expression involving δω that yields[21] $\Delta\omega^- = \xi \cdot \Delta\omega_o^-$; $\Delta\omega^+ = \xi \cdot \Delta\omega_o^+$; $\xi = 1/n_g$; $\delta\omega = \Delta\omega^- - \Delta\omega^+ = \delta\omega_o \cdot \xi$, where we define an enhancement factor $\xi \equiv 1/n_g$. For a medium that exhibits slow-light, $n_g \gg 1$, the result in eqn. 11 implies a reduction in rotational sensitivity. On the other hand, it is equally possible to achieve a condition where $0 < n_g \ll 1$, characteristic of the anomalous dispersion ($\partial n/\partial \omega < 0$) that leads to superluminal pulse propagation in the medium (of course, without violating SR)[4].

One way this process can be understood physically is in terms of an effective positive feedback (later on, we discuss another physical way to interpret this effect.) Consider, for example, the frequency shift in the CCW direction, $\Delta\omega^-$, given by $\delta\omega_o/2$ in the absence of dispersion. Imagine that the anomalous dispersion is now turned on. As the frequency of the CCW beam is moved to find the new resonance condition, it will see an index that is smaller than $n_o$. This leads to an increase in the value of $\Delta\omega^-$ needed for finding the resonance. As $\Delta\omega^-$ is increased further, the CCW beam experiences an even smaller value of the index. This positive feedback type process can be represented by an equation of the form $\Delta\omega^- = \delta\omega_o/2 + G * \Delta\omega^-$, where a simple analysis shows that the effective "feedback loop gain" G is given by $-(\omega_o/n_o)(\partial n/\partial \omega) = (1-n_g/n_o)$, which is positive for anomalous dispersion. The steady-state solution of this system is $\Delta\omega^- = [\delta\omega_o/2]*[1/(1-G)]$. Exactly the same process leads to an increase in the frequency shift of the CW beam, so that the net frequency difference is $\delta\omega_o*[1/(1-G)]$. The CAD condition ($n_g=0$, which implies $n' = -1/\omega_o$) corresponds to the feedback gain approaching unity ($G \to 1$).

However, the enhancement factor ξ diverges as we approach this limit. This is because we have neglected higher order effects. In order to take the higher order effects into account, it is necessary to consider an explicit model of a medium that can be used to realize the requisite negative dispersion. We now consider the negative dispersion associated with an absorptive resonance in a material medium that can be described in simple mathematical form for our analysis. Here, one should keep in mind that it is not feasible to use this kind of medium with undesirable intra-cavity absorption in actual experiments. Instead, one can consider using bi-frequency Raman gain in real experiment to produce similar dispersion effect without absorption[4]. However, the conclusions drawn from our discussion using an absorptive resonance model remain equally valid for the Raman gain. The expression for the dispersive index $n(\omega)$ associated with an absorptive resonance in a two-level system can be modeled as

$$n(\omega) = 1 - (A\Gamma).(\omega-\omega_o)/[\Gamma^2 + (\omega-\omega_o)^2] \qquad (12)$$

where $2\Gamma$ corresponds to the linewidth (FWHM) of the absorptive resonance. The magnitude of "$A$" corresponds to the peak value of the imaginary part of the susceptibility i.e. $A = (\chi_I)_{\omega=\omega_o}$. Thus, the value of the parameter "$A$" depends on medium properties such as the density and the strength of the resonant excitation. For our analysis, we can assume the above form of $n(\omega)$ and express it as a Taylor expansion around $\omega=\omega_o$, containing higher-order terms only up to $(\Delta\omega)^3$:

$$n(\omega) = 1 + n_1.\Delta\omega + n_3.(\Delta\omega)^3, \quad n_1 = \left.\frac{dn}{d\omega}\right|_{(\omega=\omega_o)}, \quad n_3 = \frac{1}{6}.\left.\frac{dn}{d\omega}\right|_{(\omega=\omega_o)} \qquad (13)$$

Note that in this model $n_o=1$, and $n_2=0$, since $n(\omega)$ in eqn. 12 is anti-symmetric around $\omega=\omega_o$. The dispersion coefficients $n_1$ and $n_3$ can be calculated by evaluating the derivative of $n(\omega)$ at $\omega=\omega_o$ and are given by $n_1 = -A/\Gamma$ and $n_3 = -n_1/\Gamma^2$. Substituting the series form of $n(\omega)$ given by eqn. 13 in eqn. 10, and solving eqn. 9, we again obtain a self consistent solution for the enhanced frequency splitting, for the case of $n_g=0$ or $n_1=-1/\omega_o$, to be given by:

$$\Delta\omega^- = \eta\bullet\Delta\omega_o^-; \quad \Delta\omega^- = \eta\bullet\Delta\omega_o^+; \quad \delta\omega = \Delta\omega^- - \Delta\omega^- = \eta\bullet\delta\omega_o; \quad \eta = [4\Gamma/\delta\omega_o]^{2/3} \qquad (14)$$

The enhancement factor, $\eta$, is non-linear. It does not diverge for finite values of $\delta\omega_o$, and saturates to a value close to unity as $\delta\omega_o$ approaches the extrema of the dispersion[22].

# 4. Effective Equivalence Between Rotation Sensing and Measurement of Mirror Displacement using a Ring Resonator

The derivation of these results so far has been perhaps less than transparent, especially given the presence of relativistic addition of velocities, and the apparent complexity of the Sagnac effect. As such, the physical interpretation of these results may not be obvious. Therefore, we now consider an analogous system where physical intuitions should be more easily applicable. Specifically, we establish here a simple analogy between the rotational frequency shift in a PRCG to the frequency shift in cavity resonance caused by the mirror displacement in a cavity. The later process can be considered to be a device where the frequency shift is used to measure the mirror displacement caused by a perturbation. For simplicity, let us also assume, as we have done above, that the cavity is entirely filled with a dispersive medium. The resonance condition for the cavity with a free-space length $L$ is given as

$$L = m\lambda = 2\pi m c_o / [n(\omega_o)\omega_o] \tag{15}$$

where $\omega_o$ is the cavity resonance frequency, $n(\omega)$ characterizes the dispersion due to the medium and $m$ is a large integer number. In the absence of intracavity dispersion, one can easily calculate the shift in the cavity resonance frequency for a small change $\Delta L$ in length for an empty cavity, which is given by $\Delta\omega_{ec} = -\Delta L \cdot \omega_o / L$. It is instructive to compare this expression with the first expression in eqn. (9), for example, which can be rewritten as:

$$\Delta\omega_o^- = [\omega_o/(C_o n_o)] \cdot [2\Omega A / P] = -[\omega_o / L_{eff}] \cdot \delta L_{eff}; \quad L_{eff} \equiv P,$$
$$-\delta L_{eff} \equiv P \cdot \Omega R /(n_o C_o) = L_{eff} \cdot v /(n_o C_o) \tag{16}$$

Thus, within a proportionality constant, the effect of rotation can be equated to an effective change in the length of the cavity. Therefore, it is possible to interpret the mechanism behind the enhancement in a simpler fashion by considering the actual change of length in a PRC, as modeled by eqn. 15 above.

In the presence of a change in the cavity length, $\Delta L$, and assuming that the index is dispersive, we can write the generalized form of eqn. 15 as

$$L + \Delta L = m\lambda = 2\pi m c_o / [n(\omega_o + \Delta\omega) \cdot (\omega_o + \Delta\omega)] \tag{17}$$

where $\omega = \omega_o + \Delta\omega$ is the new cavity resonance frequency due to the change $\Delta L$ in the cavity length. If the index is assumed to be of the form $n(\omega_o + \Delta\omega) = 1 + n_1 \Delta\omega$, corresponding to linear dispersion, one can easily find a solution to the above equation to determine the frequency shift $\Delta\omega_{dis}$ in terms of the empty cavity frequency shift $\Delta\omega_{ec}$ as

$$\Delta\omega_{dis} = \Delta\omega_{ec} / n_g, \quad n_g = 1 + n_1 \cdot \omega_o \tag{18}$$

where $n_g$ corresponds to the value of the group index at $\omega=\omega_o$. Note that here we have assumed the mean refractive index $n(\omega_o)$ of the medium to be unity. Eqn. (18) suggests that at extremely small value of $n_g$, for example $n_g = 0.001$, $\Delta\omega_{dis}$ can be $10^3$ times larger than $\Delta\omega_{ec}$.

This is essentially[23] the same result as what was derived above for the PRCG, if we set $n_o=1$. In the current case, however, the explanation for the enhancement is very simple. Due to the linear dispersion in the cavity, the wavelength $\lambda$ inside the medium is changing very slowly with the frequency $\omega$. Thus, for a given change in cavity length, $\Delta L$, one has to change the frequency $\omega$ by a significantly larger amount (compared to the empty cavity case) in order to ensure that the new length matches an integer number of wavelengths, thus establishing resonance. Therefore, $\Delta\omega_{dis}$ has to be much larger than $\Delta\omega_{ec}$. If this argument is extended to the case when $n_g = 0$, which occurs when the dispersion slope $n_1 = -1/\omega_o$, the wavelength inside the cavity becomes completely insensitive to any change in $\omega$. This implies that $\Delta\omega_{dis}$ has to be "infinitely large" in order to compensate for any cavity length change $\Delta L$. In other words, the cavity cannot compensate for the length change by any finite change in the field frequency. This, obviously, corresponds to an unphysical situation since the linear dispersion model can not exist over an infinite bandwidth for a true material medium. In reality, for any real physical medium, one has to consider linear dispersion over a small (or restricted) frequency range and therefore, include the higher-order dispersion effects in order to predict what exactly happens for $n_g = 0$. Using the explicit model summarized in eqns. 12 and 13, we obtain $\Delta\omega_{dis}(1 + n_1\omega_o) + n_3 \omega_o (\Delta\omega_{dis})^3 = \Delta\omega_{ec}$. The solution to the above equation for the case $n_g = 0$ or $n_1 = -1/\omega_o$ yields

$$\Delta\omega_{dis} = \eta \cdot \Delta\omega_{ec} = [2\Gamma / \Delta\omega_{ec}]^{2/3} \cdot \Delta\omega_{ec} \tag{19}$$

This suggests that at $n_g = 0$, the shift in cavity resonance $\Delta\omega_{dis}$ for the change $\Delta L$ is finite and enhanced by a factor $\eta$ with respect to the empty cavity shift $\Delta\omega_{ec}$. This result is essentially[24] the same as what is presented for the PRCG in eqn. 14. A large enhancement factor can be observed under the condition $\Gamma \gg \Delta\omega_{ec}$, which can be satisfied experimentally for extremely small values of $\Delta\omega_{ec}$ (or $\Delta L$) caused due length change (or inertial rotation for the PRCG or the RLG). As mentioned after eqn. 18, note that $\eta$ has a nonlinear dependence on $\Delta\omega_{ec}$. Specifically, $\eta$ decreases with increasing $\Delta\omega_{ec}$, approaching a value of the order of unity as the frequency shift becomes comparable to the width of the dispersion profile.

A major feature of the dispersion condition $n_g = 0$ is that the enhanced shift is also accompanied by an almost equally large broadening of the cavity resonance linewidth. Before we consider the effect of this broadening on the ability to measure very small length change (or, equivalently, the ability of the PRCG or the RLG to measure very small rotations), it is instructive first to discuss the mechanism behind this broadening. To this end, let us consider first the situation when $\Delta L = 0$. The linewidth of the cavity can be calculated by considering the dephasing $\Psi$ experienced by the light beam for a single pass through the resonator, under the condition where $\omega \neq \omega_o$. It is easy to show that this is given by $\Psi = n(\omega)\omega L/c_o - \omega_o L/c_o$. It is also easy to show that if $\gamma$ represents the linewidth (FWHM) of a cavity, the half-width frequency ($\omega_o + \gamma/2$) will correspond to a value of $\Psi = \pi/\mathbb{F}$, where $\mathbb{F}$ is the finesse of the cavity. Using this fact and assuming again a simpler linear dispersion model i.e. $n(\omega) = 1 + n_1 \cdot \Delta\omega$, one can obtain the linewidth of the dispersive cavity as $\gamma_{dis} = \gamma_{ec}/n_g$, where $\gamma_{ec} = 2\pi c_o/(L\mathbb{F})$ is the empty cavity linewidth in the absence of dispersion. This shows that, just like the frequency shift needed to compensate for a change in the cavity length, the linewidth of the cavity also depends inversely on $n_g$. Again, for a small value of $n_g = 0.001$, for example, $\gamma_{dis}$ will be a thousand times broader than $\gamma_{ec}$. This can be understood by the same physical argument as earlier, in terms of the insensitivity of the wavelength as a function of frequency under this particular dispersion condition. Therefore, it suggests that the cavity can be nearly resonant over a large range of frequencies. For an ideal linear dispersion medium exactly at $n_g = 0$, the cavity will resonate at all frequencies, since the medium wavelength is now completely independent of the frequency. Of course, this will imply an unphysical, infinitely large linewidth for the cavity. One can extend the analysis for a realistic situation by considering a medium with a limited range of linear dispersion. Choosing again the explicit model summarized in eqns. 17 and 18, we find $\gamma_{dis} = \gamma_{ec}/[n_g + n_3 \omega_o (\gamma_{dis})^2]$. When $n_g = 0$, the broadened cavity linewidth is given by $\gamma_{dis} = [\Gamma^2 \cdot \gamma_{ec}]^{1/3}$. This condition is also referred to as the white-light cavity. Note, however, that while the cavity linewidth is broadened due to the negative dispersion, *the cavity decay rate (and, therefore the quality factor and the buildup factor) remains unchanged*[25,26,27]. One can similarly estimate the linewidth for the case of $\Delta L \neq 0$ to be $\gamma_{dis} = \gamma_{ec}/n_g\big|_{\omega=\omega_o+\Delta\omega_{ec}} \simeq (\eta/3)\cdot\gamma_{ec}$, where $\eta$ is defined in eqn. 19. This equation shows that for $\Delta L \neq 0$, the shifted, dispersion-enhanced cavity linewidth is inversely proportional to the *local* value of the group index: $n_g\big|_{\omega=\omega_o+\Delta\omega_{ec}}$ and therefore, will get narrower as the shifted resonance moves further away from $\omega=\omega_o$ for increasing $\Delta L$.

The analysis presented here involves minor approximations. In order to confirm the validity of these approximations, we have also simulated the behavior of the resonator in the presence of a dispersive medium. Figure 4 shows the enhancement factor, $\eta$, as a function of the empty-cavity frequency shifts needed to compensate for a real or effective (rotation induced) change in the cavity length. As can be seen, the simulation agrees well with the analytical estimate. Figure 5 shows explicitly a representative case, where the frequency shift in the absence of the dispersive medium is only 300 kHz, while the enhanced shift is about 9.5 MHz.

Note that in real experiments, it may be necessary to fill only a part of the cavity with the dispersive medium. For this scenario, one can carry out a similar analysis, considering the medium length $\ell$ to be smaller than the cavity length L. In this case, we find all our conclusions reached here to remain valid, except that the dispersion condition $n_g = 0$ for linewidth broadening and sensitivity enhancement gets modified to a value of group index corresponding to $n_g = 1 - L/\ell$.

## 5. Counterbalancing Effects of Enhancement in Sensitivity and Linewidth Broadening in a Fast-Light Enhanced Ring Resonator

Consider now the situation where our goal would be to measure the change in the cavity length $\Delta L$ caused due any physical process or due effectively to a rotation. We now explicitly show how the cavity linewidth broadening in the case of a dispersive cavity affects our ability to measure ΔL with a higher sensitivity η, predicted earlier at $n_g = 0$. The change ΔL is indirectly determined by measuring the shift in the cavity resonance frequency Δω. For the case of no dispersion, the minimum measurable change in length $[\Delta L]_{min}$ is related to minimum measurable $[\Delta \omega]_{min}$ as

$$[\Delta L_{ec}]_{min} = [\Delta \omega]_{min} \cdot (L/\omega_o) \tag{20}$$

where $[\Delta \omega]_{min}$ for an empty cavity is given by:

$$[\Delta \omega]_{min} = \gamma_{ec} / SNR \tag{21}$$

Our previous discussion also suggests that the minimum measurable change in length for the dispersive cavity is

$$[\Delta L_{dis}]_{min} = ([\Delta \omega_{dis}]_{min} / \eta) \cdot (L/\omega_o) \tag{22}$$

For the dispersive cavity with a broadened linewidth $[\Delta \omega_{dis}]_{min}$ is given by

$$[\Delta \omega_{dis}]_{min} = \gamma_{dis}\big|_{\omega = \omega_o + \Delta \omega_{ec}} / SNR = (\eta/3)(\gamma_{ec}/SNR) \tag{23}$$

Substituting eqn. 23 in eqn 22, and eqn, 21 in eqn. 20, one finds, assuming the SNR to be the same in each case, that $[\Delta L_{dis}]_{min} = [\Delta L_{ec}]_{min}/3$. Therefore, the minimum measurable $[\Delta L]_{min}$ remains nearly unchanged for a dispersive cavity due to cavity linewidth broadening. The factor of 3 improvement is perhaps not significant enough to warrant the additional complications and sources of drifts and noise that may come into play when the dispersive element is introduced.

## 6. True Reduction in the Smallest Measurable Rotation Rate using a Fast-Light Enhanced Ring Laser Gyroscope

The conclusion reached at the end of the previous section is drastically altered if we consider a situation where instead of the cavity being fed by an external laser, the dispersive cavity itself contains an active gain medium. Let us also assume that the gain spectrum associated with the gain medium is considerably broad compared to the cavity resonance, so that the effect of the dispersion due to the gain medium itself on the cavity response can be ignored. In the context of rotation measurement, such a cavity already exists in the RLG. Of course, the RLG can be also used to measure an actual change in the length of the cavity. In simple terms, as the cavity resonance frequency shifts due to $\Delta L$ (actual, or effective due to rotation), the lasing process inside the cavity will move to the new cavity resonance frequency. As a result, one can directly measure the frequency shift $\Delta\omega$ corresponding to the $\Delta L$ by simply beating the laser beam outside the cavity with a reference laser beam. In the case of an RLG, the beat between the CW and the CCW beams outside the cavity gives a direct measure of the frequency shift which is proportional to the rotation rate. The minimum measurable $\Delta L$ i.e. $[\Delta L]_{min}$ for the RLG is decided by the measured linewidth of the laser beam[17,28]. The fundamental quantum noise limited linewidth of the laser in such a case is given by[28] $(\Delta\omega)_{laser} = [1/\tau_c]/\sqrt{P_{out} \cdot \tau_m / \hbar\omega}$, where $\tau_c$ is the decay time of the photon (or field) inside the cavity, which is the inverse of the empty cavity linewidth: $\gamma_{ec} = 1/\tau_c$, and we have assumed a unity quantum efficiency for the detector. Here, $P_{out}$ is the output power and $\tau_m$ is the measurement time. The quantity $\sqrt{P_{out} \cdot \tau_m / \hbar\omega}$ in the denominator simply corresponds to the square root of the number of photons observed during the measurement time $\tau_m$. For a laser in a coherent state, this is therefore the uncertainty $\Delta n$ in the number of photons measured. Note further that $\Delta\omega \cdot \tau_c$ represents the uncertainty $\Delta\phi$ in the phase of the laser field, accumulated over its lifetime inside the cavity. As such, this expression for laser linewidth can simply be interpreted as a manifestation of the number-phase uncertainty relation: $\Delta\phi \cdot \Delta n = 1$. Other derivations of the laser linewidth also reduce to this expression when the measurement time is taken into account[29,30,31,32,33]. Given this expression for the laser linewidth, one finds $[\Delta L]_{min}$ to be

$$[\Delta L_{ec}^{RLG}]_{min} = [\Delta\omega]_{min} \cdot L/\omega_o = (\Delta\omega)_{laser} \cdot L/\omega_o = [(1/\tau_c)/\sqrt{P_{out} \cdot \tau_m / \hbar\omega}] L/\omega_o \qquad (24)$$

For comparison, consider a situation where one uses a PRCG instead of an RLG, with the output power of the cavity being the same as the output power of the RLG. It can be easily shown that $[\Delta L]_{min}$ for an active cavity shown here is the same[17,28] as the one for a passive cavity: $[\Delta L_{ec}^{passive}]_{min} = [\Delta\omega_{ec}]_{min} \cdot L/\omega_o = (\gamma_{ec}/\sqrt{N}) \cdot L/\omega_o = [\Delta L_{ec}^{RLG}]_{min}$, where N is the number of photons observed during the measurement time in each case, with unity quantum efficiency.

In the case of a cavity loaded with the negative dispersive medium, although $\gamma_{dis}$ at $n_g = 0$ is much broader than $\gamma_{ec}$, it is important to note that the decay time of the field inside the dispersive cavity remains unchanged and is determined purely by the external properties of the cavity such as losses and mirror reflections, etc. Therefore, if one constructs an RLG loaded with an additional negative dispersive medium, as described above, the quantum-noise limited laser linewidth $(\Delta\omega)_{laser}$ will remain unchanged. In this case, the minimum measurable $\Delta L$ is found to be $[\Delta L_{dis}^{RLG}]_{min} = [\Delta L_{ec}^{RLG}]_{min}/\eta = [\Delta L_{ec}^{passive}]_{min}/\eta$. This represents a real improvement in sensitivity of measurement by a factor $\eta$. Therefore, equivalently, the minimum measurable rotation is given by $[\Delta\Omega_{dis}^{RLG}]_{min} = [\Delta\Omega_{ec}^{RLG}]_{min}/\eta = [\Delta\Omega_{ec}^{passive}]_{min}/\eta$. Thus, the sensitivity in rotation measurement can be improved by using an active gain medium inside the dispersive cavity. For proper choice of parameters, the enhancement in sensitivity $\eta = [2\Gamma/\Delta\omega_{ec}]^{2/3}$ can be made very large.

## 7. Experimental Considerations, Numerical Estimates and the Prospect of Measuring the Lens-Thirring Rotation

Experimentally, a fast-light enhanced ring laser gyroscope may be realized as follows. Consider a ring cavity incorporating a broadband gain medium such as a Ti:Sapphire crystal pumped by an Argon laser. We also assume that it incorporates mode-selective elements such as a bi-refringent filter and an etalon, so that the laser operates in only two degenerate, counter-propagating longitudinal modes. Such a device would work as a ring laser gyroscope; in the presence of rotation perpendicular to the cavity plane. We assume that the operating frequency is tuned close to the D2 transition in $^{85}$Rb. We now place a vapor cell containing $^{85}$Rb inside this cavity, with anti-reflection coated windows. The vapor is optically pumped to produce a population inversion between the F=2 and F=3 hyperfine ground states. A dual-frequency Raman pump, detuned from the P manifold, is the applied. In the presence of this pump, a probe beam will experience two gain peaks, each corresponding to the condition where the probe is two-photon resonant with one of the pump frequencies. This is precisely the scheme that was employed in demonstrating loss-free generation of fast-light[4]. The optical pumping beam as well as the dual-frequency pump beams will be inserted into the cavity using a polarizing beam splitter (PBS), and extracted from the cavity using another PBS. The lasing modes inside the cavity will play the role of the probe beam, with its frequency chosen to be at the center of the two gain peaks, so that it experiences the steepest negative dispersion.

Recently, in an experiment the details of which are being reported elsewhere,[34] we demonstrated the key aspects of such a device. In particular, we inserted a Rb vapor cell in a passive ring cavity (without the Ti:Sapphire gain medium), and generated negative dispersion using the dual-pumped Raman gain outlined above. An external probe was made to resonate in the cavity, and its behavior was studied with and without activating the Raman gain doublet induced negative dispersion. As expected from the discussions presented above, the cavity displayed the White Light behavior: the resonance linewidth of the probe became broadened in the presence of the dispersion.

In this experiment, it was difficult to demonstrate the enhanced sensitivity directly for the following reason. The enhancement factor is non-linear: it decreases with increasing values of the empty-cavity frequency shift, $\Delta\omega_o$ (corresponding to $\Delta L$, or equivalently, a rotation rate). Furthermore, in order for the enhancement to be evident, the value of the loaded-cavity frequency shift, $\Delta\omega_o'$ (i.e., the enhanced shift) must be less than the dispersion bandwidth. Thus, for the limited dispersion bandwidth realized in this experiment, one must use a very small value of $\Delta L$ in order to observe an enhancement. This in turn requires the use of a resonator that has a much higher finesse than the one used in ref., and a more precise voltage supply for the PZT. The required modifications are non-trivial, and efforts are underway in our laboratory to implement these changes to the apparatus.

Given this constraint, we modified the experimental configuration in order to demonstrate the enhancement effect indirectly[35]. As can be seen from the theory presented above, the modification factor of the sensitivity depends on the inverse of the group index, independent of its value. For negative dispersion corresponding to a nearly null value of the group index, the factor is greater than unity, and corresponds to enhancement of sensitivity. For positive dispersion, the factor is less than unity, corresponding to reduction in sensitivity. While the former is hard to observe using our current apparatus, the later is much easier to demonstrate. This is due to two reasons. First, the modification factor can be tuned to be close to unity by controlling the slope of the positive dispersion. Second, since the loaded-cavity frequency shift is less than the empty cavity frequency shift, it is easy to keep its value within the dispersion bandwidth. We demonstrated this reduction in sensitivity by using two alternative modification of the scheme for exciting the vapor cell. In one version, we simply replaced the dual-frequency Raman pump with a single frequency pump, and tuned the probe to be at the center of the corresponding Raman gain. In another version, we turned off the optical pumping beam, thus eliminating Raman gain, and tuned both the (single-frequency) pump and the probe to resonance with the P-manifold, while maintaining the two-photon resonance condition. This produced an electro-magnetically induced transparency (EIT) for the probe. In either case, the probe experiences positive dispersion corresponding to slow light. As expected, we observed reduction in sensitivity proportional to the inverse of the group index, which was measured independently. This experiment confirmed that the sensitivity modification varies in a manner that agrees with our theory, thus establishing confidence in the claim that negative dispersion near the CAD condition will yield enhancement.

In most experiments involving EIT in a vapor cell, it is impossible to achieve a very high degree of transparency, primarily because of uncontrolled optical pumping into Zeeman sublevels. As such, there is always some absorption present, and experiments purporting to make use of the EIT process suffers from deleterious effects of this absorptiob. The situation for the doubly-pumped Raman gain system is quite different. With proper choice of parameters, as evidenced in references 4 and 34, the probe experiences true gain. Furthermore, since the probe is positioned in-between the two gain peaks, the gain seen by it can be tuned very close to zero if desired, or adjusted to compensate for residual losses from the vapor cell windows and the PBS's. Thus, the role of the intra-cavity medium can be tailored to be to produce purely the dispersion, without any deleterious effect of any residual loss or gain.

One may also think that since the cavity becomes very sensitive to rotation near the CAD condition, any noise in the system is also likely to be amplified, thus offsetting any real advantage. However, this is not the case. To see why, note first that in designing any apparatus for ultimate performance, stabilization schemes have to be in place in order to reduce the systematic noise sources to a level so that one is limited only by quantum noise. For the fast-light enhanced gyroscope, the same rule would apply. The enhanced sensitivity to systematic sources of noise in this case simply means that the stabilization schemes have to be more robust. Since the fundamental quantum noise, which affects the linewidth of the ring laser, is not affected by the fast-light enhancement of sensitivity, as discussed earlier, it is possible to achieve true enhancement in the capability for measuring very small rotations.

As a numerical example, let us consider a table-top size RLG, with a value of P/A= $2m^{-1}$ (For an equivalent circular shape, the radius is 1 meter). Let us assume that the operating mean frequency is about $5\times10^{14}$ Hz, the cavity finesse is about $10^3$, and the output power is 1 mW. A quick estimate then shows that in the absence of fast-light enhancement, the minimum measurable rotation rate for an observation time of 1 sec is about $1.5\times10^{-5}$ $\Omega_\oplus$, where $\Omega_\oplus$ is the Earth rotation rate. Suppose now we insert a negative dispersion medium in the cavity, filling only a part l of the path. As shown in detail in ref **Error! Bookmark not defined.**, this can still yield the same degree of enhancement as formulated above, as long as the value of $n_g$ is chosen to equal (1-P/l). For a realistic dispersion linewidth of $\Gamma=2\pi*10^6$ $sec^{-1}$, the enhancement factor corresponding to the minimum measurable frequency shift (without the dispersion enhancement) is about $1.8*10^6$, and the minimum measurable rotation rate then becomes about $10^{-11}$ $\Omega_\oplus$, for an observation time of 1 sec. Of course, this number can be improved further by increasing the observation time, or by increasing the area of the RLG. Note that this number is already smaller than the Lens-Thirring rotation rate of $5.6*10^{-10}$ $\Omega_\oplus$ expected on an earth-bound experiment, according to Einstein's theory of gravity. This resolution should also therefore be high enough to distinguish between the predictions made by different theories of gravity (Einsteins' Theory, the Brans-Dicke-Jordan Theory, and the Ni Thery) for the magnitude of the Lens-Thirring effect[16,36,37].

## 8. Conclusion and Acknowledgments

To summarize, we have described a ring resonator based optical gyroscope whose sensitivity for measuring *absolute* rotation is enhanced via the presence of an intra-cavity dispersive medium. Specifically, we show that the enhancement occurs if the dispersion is anomalous, characteristic of superluminal light propagation. Under an idealized model where the index varies linearly over all frequencies, the enhancement factor is given by the inverse of the group index, and is maximum when the group index is null, corresponding to the so-called Critically Anomalous Dispersion (CAD) condition (i.e., where the group velocity becomes infinite). For a realistic medium, the anomalous dispersion has a limited bandwidth. When this constraint is taken into account, the divergence at the CAD condition is eliminated, while still yielding a very large enhancement. For realistic conditions, the enhancement factor is as high as $10^6$. We also show how the offsetting effect of the concomitant broadening of the resonator linewidth can be circumvented by using an active cavity. In addition to this fast-light based effect, we have shown how normal dispersion used for slow light can enhance *relative* rotation sensing in a specially designed Sagnac interferometer. In this case, the enhancement is given by the slowing factor.

We acknowledge useful discussions with Prof. S. Ezekiel of MIT. This work was supported in part by the Hewlett-Packard Co. through DARPA and the Air Force Office of Scientific Research under AFOSR contract no. FA9550-05-C-0017, and by AFOSR Grant Number FA9550-04-1-0189.

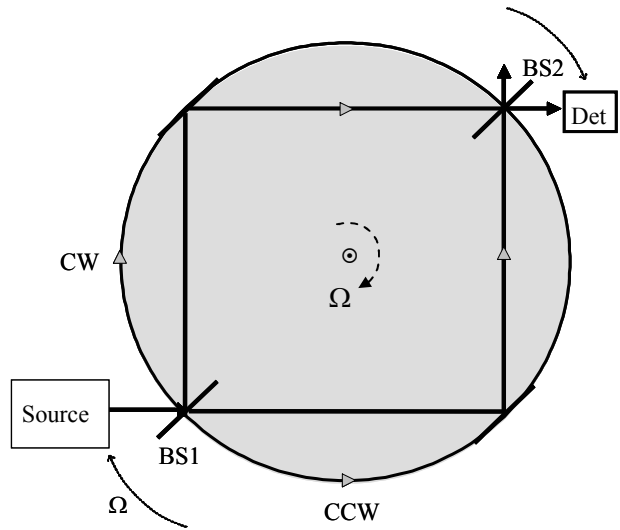

**Figure 1** Schematic illustration of a Mach-Zehnder type Sagnac interferometer with counter-propagating beams in a circular optical path (the actual paths shown are rectilinear for simplicity).

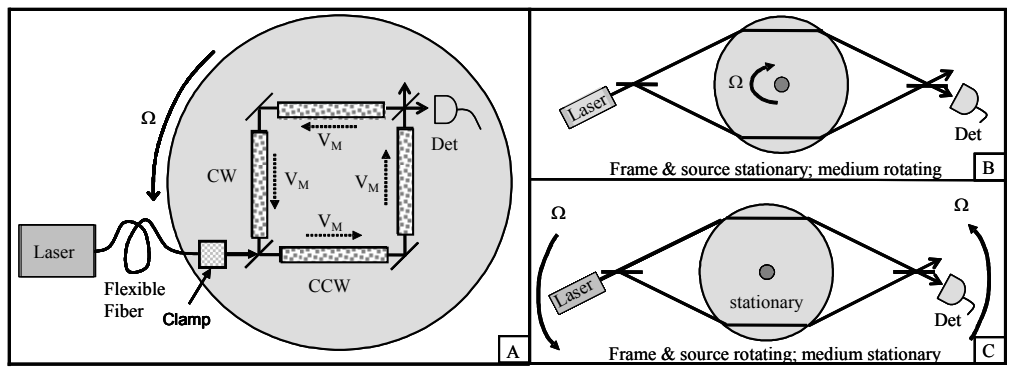

**Figure 2** Interferometric setup for rotation sensing using Doppler effect (a) generalized case corresponding to uniform medium translation (b) stationary source and frame, rotating medium (c) co-rotating source and frame, stationary medium

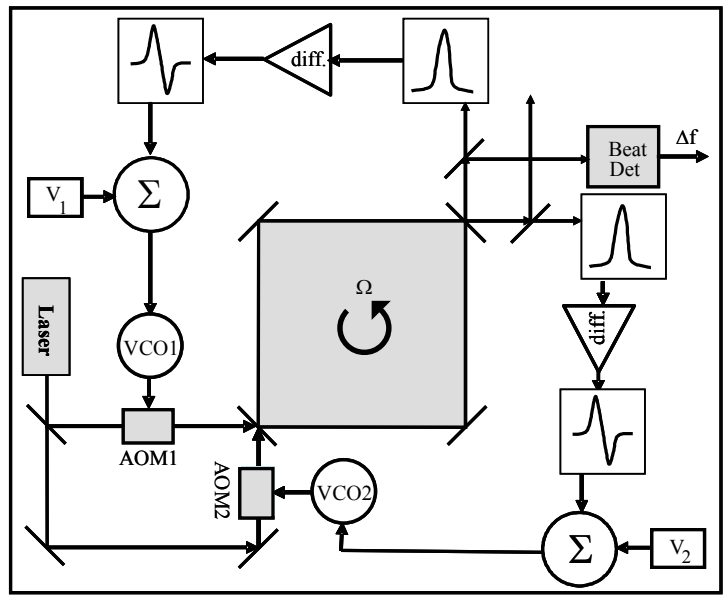

**Figure 3** Schematic diagram of a passive resonator gyroscope showing detection mode for beat frequency measurement and active closed-loop control for compensating random drift

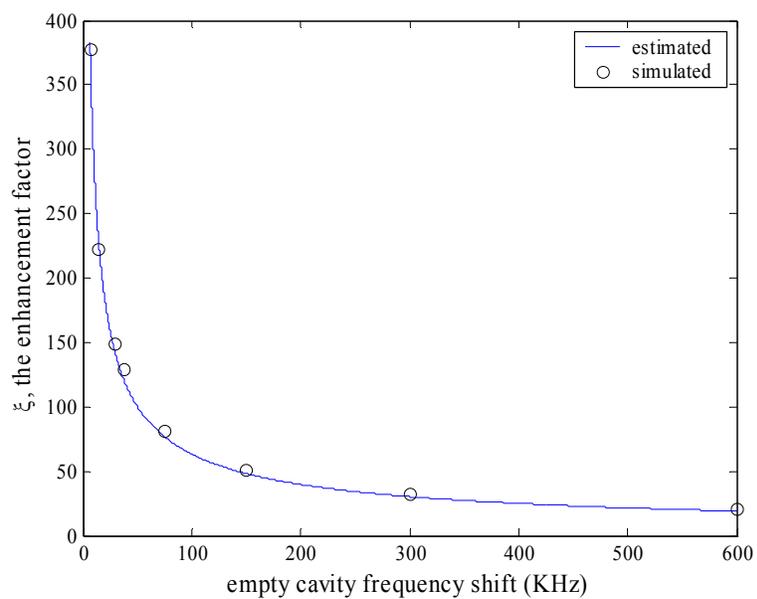

**Figure 4:** Comparison of the analytical estimates and the numerical simulations of the enhancement factor.

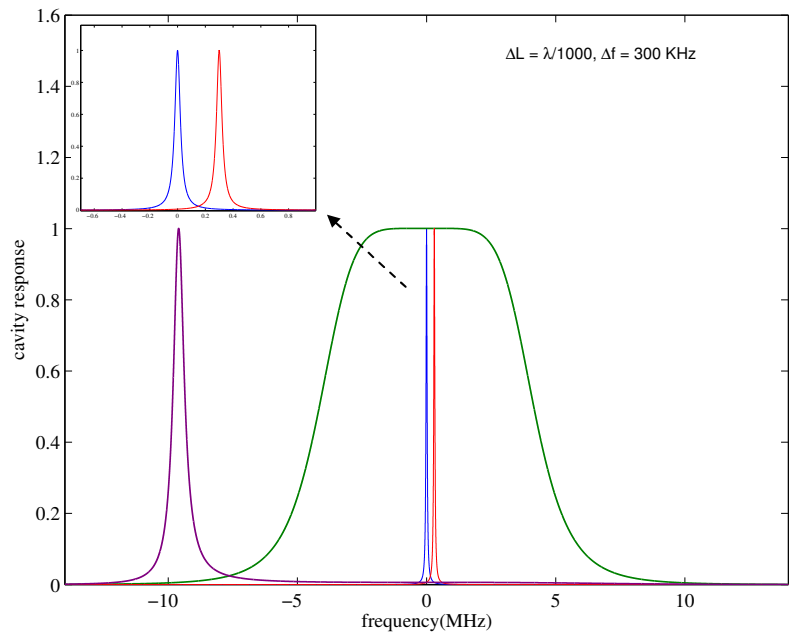

**Figure 5:** Illustration of enhancement of frequency shift due to negative dispersion.

[21] More exactly, the value of $\xi$ is slightly different for the CCW and the CW cases. In the CCW case, it is modified by a factor of $n_o/n(\omega_o + \Delta\omega^-)$, and in the CW case, it is modified by a factor of $n_o/n(\omega_o + \Delta\omega^+)$. However, in a realistic dispersive medium, especially under the CAD condition, these factors remain very close to unity.

[22] Again, more exactly, the value of $\eta$ is slightly different for the CCW and the CW cases. In the CCW case, it is modified by a factor of $n_o/n(\omega_o + \Delta\omega^-)$, and in the CW case, it is modified by a factor of $n_o/n(\omega_o + \Delta\omega^+)$. However, in a realistic dispersive medium, especially under the CAD condition, these factors remain very close to unity.

[23] Recall that in the case of the PRCG, the enhancement factor is slightly different for the CCW and the CW case, as summarized in footnotes 21 and 22, by factors that are determined by the values of the index at resonance. The difference between the model where there is actual change in the length and the case of rotation-induced effective length change is also embodied in these factors. This difference arises from the fact that the effective values of a length change corresponding to an actual rotation depends inversely on the value of the index, as can be seen from eqn. 20. However, since these factors remain very close to unity (with a maximum deviation of the order of $10^{-7}$ in a typical medium such as atomic vapor, near the CAD condition), the analogy is accurate for all practical purposes.

[24] Note that the expression for $\eta$ in eqn. 14 can be expressed as $[2\Gamma/(\delta\omega_0/2)]^{2/3}$. The expression $\Delta\omega_{ec}$ in eqn. 19 corresponds to $(\delta\omega_0/2)$ in eqn. 14.